\documentclass[epsf,preprint,aps]{revtex4}
\usepackage{graphicx}
\usepackage{bm}

\date{\today}

\begin{document}

\title{To the positive miscut influence \\ on the crystal collimation efficiency}

\author{V.~V.~Tikhomirov, A.~I.~Sytov}
\email{vvtikh@mail.ru, alex_sytov@mail.ru} \affiliation {Research
Institute for Nuclear Problems, Belarus State University,
Bobruiskaya 11, 220030 Minsk, Belarus}

\begin{abstract}
The paper concerns the crystal based collimation suggested to
upgrade the Large Hadron Collider collimation system. The issue of
collimation efficiency dependence on the muscut angle
characterizing nonparallelity of the channeling planes and crystal
surface is mainly addressed. It is shown for the first time that
even the preferable positive miscut could severely deteriorate the
channeling collimation efficiency in the crystal collimation UA9
experiment. We demonstrate that the positive miscut influence can
increase the nuclear reaction rate in the perfectly aligned
crystal collimator by a factor of 4.5. We also discuss the
possible miscut influence on the future LHC crystal collimation
system performance as well as suggest simple estimates for the
beam diffusion step, average impact parameter of particle
collisions with the collimator and angular divergence of the
colliding particle beam portion.

\end{abstract}
\maketitle

\section{Introduction}

Crystal based collimation was proposed to facilitate the beam halo
cleaning at large accelerators long ago. Its application to the
LHC upgrade becomes more and more topical \cite{as1,sca,sca2,pre}.
The basic idea is to use a bent crystal in channeling mode to
deflect halo particles by relatively large angles to high impact
parameters of particle collisions with an absorber
\cite{as1,sca,sca2,pre,els}.

\begin{figure}[b]
\hspace{-0.0cm} \vspace{0cm} {\includegraphics[width=15cm]
{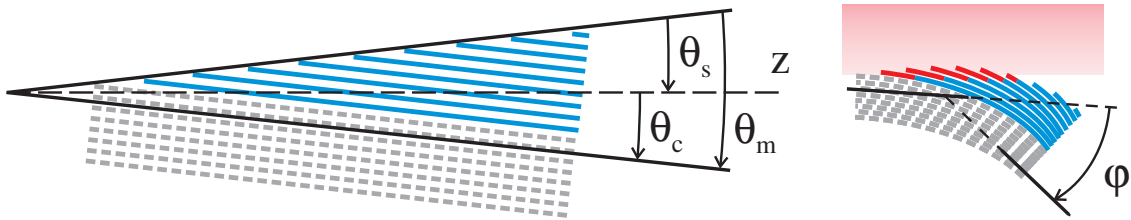}}
 \caption {A crystal with positive miscut angle before (left) and after (right)
 bending with angle $\varphi$. $\theta_c$ (positive), $\theta_m$ ({\bf positive}) and $\theta_s = \theta_c -
 \theta_m$ (negative) are, respectively, the crystal plane misalignment angle, miscat angle and
 crystal surface misorientation angle, all measured in the direction of crystal
 bending, at that the angles $\theta_c$ and $\theta_s$ --  from the $z$ axis, parallel to the
 velocity of the particle just touching the crystal, and the angle $\theta_m$ -- from the crystal surface  direction.
 Particles, moving from the left to the right with small impact parameters,
 enter the crystal through the lateral (upper) crystal surface. }
\end{figure}

If the first particle collision with the crystal collimator occurs
at sufficiently small particle incidence angle w.r.t. the crystal
planes (at pure alignment, $\theta_c=0$), the probability of
particle capture into the channeling regime reaches its maximum
\cite{sca3}. However even a small nonparallelity of the lateral
crystal surface with atomic planes, characterized by the muscut
angle $\theta_m$, is able to severely disturb the motion of
particles hitting the crystal with small impact parameters. In
particular, if the miscut angle is negative, the channeling motion
can be interrupted before the particle reaches the exit crystal
face \cite{els}. Since a considerable number of such particles
will not reach an absorber fast, the negative miscut is
recommended to be avoided \cite{els}. By this reason the positive
one (see Fig. 1) was chosen for the recent UA9 experiments
\cite{sca} aimed to demonstrate the viability of crystal
collimation. However the nuclear reaction rate in the perfectly
aligned crystal collimator about five time exceeding the
theoretically predicted \cite{sca2} value was observed. In this
paper we for the first time investigate the influence of positive
miscut on collimation efficiency and demonstrate that it gives
rise to up to 4.5 time increase of the nuclear reaction rate in
the collimator. We also predict the low influence of the positive
miscut on the efficiency of the future LHC crystal collimation
system but first suggest simple estimates for the beam diffusion
step, impact parameter of particle collisions with the collimator
and angular divergence of the part of the beam particles colliding
with the latter for the first time.

\section{Particle diffusion in the accelerator ring}

When positively charged particles strike a modestly bent crystal
moving strictly parallel to its planes, they are captured into the
regime of stable channeling motion with a probability of 80-85\%
\cite{sca3}. However if the incident particle beam possesses an
angular divergence, the "channeling probability" decreases by the
value $\Delta P_{ch} \propto \langle \vartheta^2 \rangle$, where
$\langle \vartheta^2 \rangle$ is the mean square of the incident
beam divergence angle, assumed here to be considerably smaller
than the critical channeling angle $\vartheta_{ch}$. This decrease
remains negligible only if $\langle \vartheta^2 \rangle \leq 0.01
\vartheta_{ch}^2$. Fortunately, the angular divergence of the beam
portion striking a primary collimator first time often satisfies
this rigorous condition. However if a particle has not been
captured at the first passage through the collimator, its
deflection can reach significant values. This actually pertains to
the 15-20\% of particles escaping channeling at the crystal
penetration through its normal entrance (transverse to the beam)
face. Particles entering a crystal with negative miscut \cite{els}
at small enough impact parameters are angularly dispersed even
stronger. Because of this and also since a miscut can not be
avoided in practice, the positive miscut crystal orientation is
commonly preferred \cite{els}. In this case, however, the
particles with the small enough impact parameters enter the
crystal through its lateral face. Most of them avoid capture even
in the case of pure crystal alignment. The uncaptured particles
are scattered nearly the same way as the ones in amorphous matter,
acquiring the average deflection angle squared proportional to the
length $\Delta z$ of particle path through the crystal. Besides
the angles of miscut $\theta_{m}$, crystal bending $\varphi = l/R$
and crystal plane misalignment $\theta_{c}$, the $\Delta z$ value
depends on the particle impact parameter $\Delta$ with the crystal
collimator, the value of which needs special consideration.

\begin{table*}
\caption{\label{tab:diffusionstep}Beam diffusion parameter.}
\begin{center}
\begin{tabular}{|c|c|c|c|c|c|}
\hline
Accelerator & $\varepsilon$ &~$\tau$&$\sigma(\mu m)$&$\rho_c/\sigma$&$\delta$ \\
\hline
SPS UA9                    &120GeV&10h&1010&3.5&0.086 nm\\
SPS UA9        &120GeV&4min&1010&3.5&13 nm\\
LHC                   &7TeV&10h&200&6&5.4$\mu m$\\
LHC            &7TeV&10h&420&6&11.4$\mu m$\\
\hline
\end{tabular}
\end{center}
\end{table*}

Having limited access to the parameters of particle motion in the
accelerators, not mentioning the tools allowing to simulate a
number of them, we need to suggest simple estimates for the
former. To start with, recall that particle collisions with
crystal collimator originate from the particle diffusion in the
radial beam direction caused by intra beam collisions, scattering
by residual gas, elastic scattering at the interaction point, etc.
\cite{pre}. Since a joint description of all these processes is
hardly available, we will proceed from a simple estimate based on
the accelerator beam lifetime $\tau$, particle revolution period
$T$, r.m.s. beam radius $\sigma$ and collimator radial coordinate
$\rho_c$. We will also assume, as in \cite{pre}, for simplicity,
that the beam is axially symmetric and possesses normal
distribution
\begin{equation} 
\frac{dN}{d\rho} =\frac{N \rho}{\sigma^2} \exp\left
(-\frac{\rho^2}{2 \sigma^2}\right)
\end{equation}
in particle radial number density integrated over the particle
revolution period T. Here N is a total number of particles in the
ring. Introducing a {\it diffusion step} $\delta$, an average
increase of radial coordinate acquired during one revolution by
particles reaching the collimator position, one can express the
particle loss rate in two ways:
\begin{equation} 
\frac{dN}{dt} =\frac{N}{\tau}=\left(\frac{dN}{d
\rho}\right)_{\rho_c}\frac{\delta}{T},
\end{equation}
coming directly to an estimate
\begin{equation} 
\delta =\frac{\sigma^2 T}{\tau \rho_c}\exp\left (\frac{\rho_c^2}{2
\sigma^2}\right).
\end{equation}
Table I illustrates Eq. (3) application to the cases of both the
UA9 experiment and IR7 beta collimation region at the LHC
\cite{as2}. Note that the exponential dependence on the collimator
aperture squared leads to a really drastic $\delta$ difference in
the cases of collimation dedicated UA9 experiment with low
intensity beam and $\rho_c \simeq 3.5 \sigma$ and the intensive
LHC beam and $\rho_c = 6 \sigma$.

\section{Particle impact parameter and deflection angle}

To simulate both the particle impact parameter and angular
deflection at the moment of the first collision with the
collimator we will proceed from the usual pseudoharmonic
representation
\begin{equation} 
x(\psi)= x_0 cos\psi
\end{equation}
of the betatron oscillations. Here $\psi$ and $x_0 =
\sqrt{\varepsilon \beta}$ are, respectively, the oscillation phase
and amplitude, the latter of which is determined by the beam
emittance $\varepsilon$ and accelerator beta function $\beta$.
Particle collisions with the collimator become possible since the
amplitude $x_0$ reaches the transverse collimator coordinate $x_c
= \rho_c$ (from here on we consider betatron oscillations in some
transverse plane parallel to $x$ axis). Collision really occurs if
$x(\psi) \geq x_c$ or $|\psi| \leq \psi(x_0)$, where
\begin{equation} 
\psi(x_0 )= arccos \left (\frac{x_c}{x_0}\right)  \simeq \sqrt{2
(x_0-x_c)/x_c}
\end{equation}
(we assumed that $x(\psi) - x_c \ll x_c$). The particle direction
of incidence on the crystal will be described by the angle
\begin{equation} 
\theta(\psi) = -\frac{x_0}{\beta} \left[ sin \psi
-\frac{1}{2}\frac{d \beta}{ds} cos \psi\right]
\end{equation}
of the velocity deviation from the direction of motion of the
particle just touching the collimator having $x_0 = x_c$.
\begin{figure}[b]
\hspace{-0.0cm} \vspace{0cm} {\includegraphics[width=10cm]
{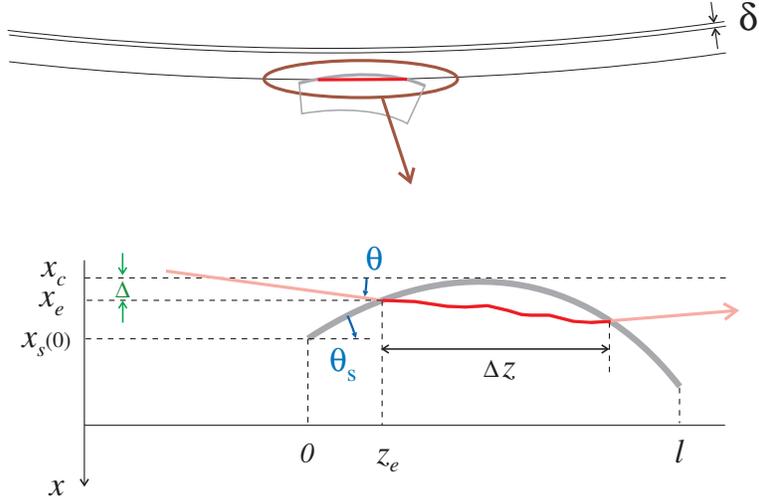}} \caption {A particle entering the crystal through the
lateral surface. The crystals extends from $z=0$ to $z=l$, $x_c$
is the collimator radial coordinate, $(x_{e}=x_s(z_{e}), z_{e})$
is the particle enter point while $(x_s(0), 0)$ is the left upper
corner of the crystal, $\theta$ and $\theta_s$ are the deflection
particle angle and crystal surface misalignment one, respectively,
$\Delta z$ is the length of particle trajectory inside the
crystal, $\delta$ and $\Delta$ are, respectively, the particle
diffusion step and impact parameter. }
\end{figure}

The process of increase of the betatron oscillation amplitude was
simulated using the formula
\begin{equation} 
x_0(n) = x_0(n-1) + 2 \delta \xi_1,
\end{equation}
where $n=1,2,..$ is the number of particle revolution in the ring
after the moment when $x_0(0)= x_c$. $\xi_1$ (as well as $\xi_2$
and $\xi_3$ below) are random numbers uniformly distributed
through the interval $(0, 1)$. Note that in average $\langle
x_0(n) \rangle = \langle x_0(n-1)\rangle + \delta $, in agreement
with the $\delta $ definition.

\begin{figure}[t]
\hspace{-0.0cm} \vspace{0cm} {\includegraphics[width=8.0cm]
{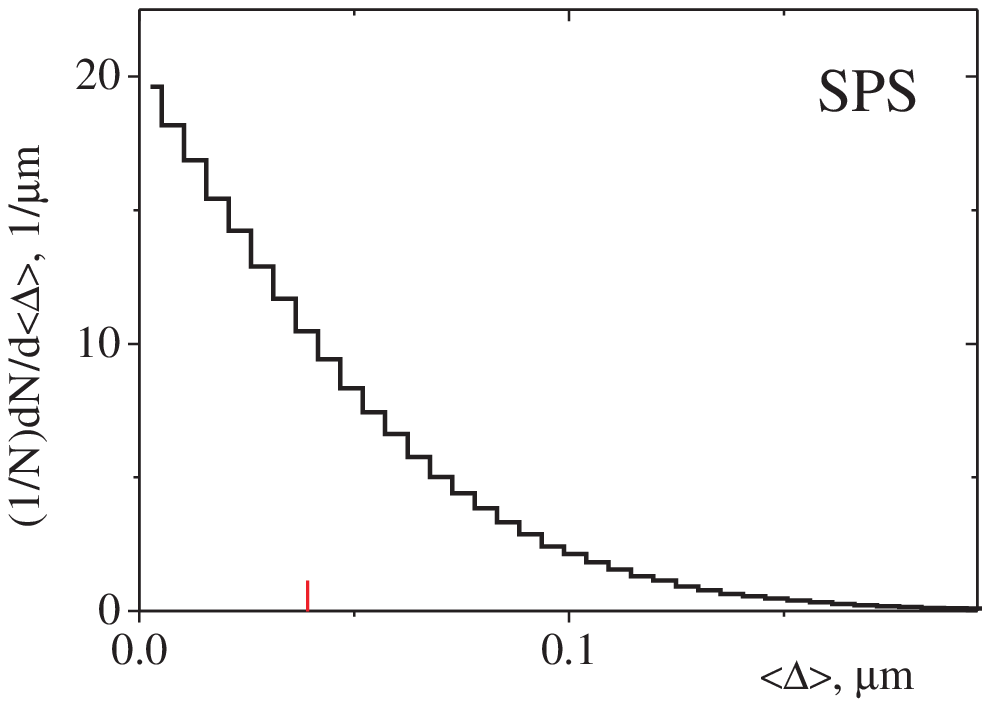}}
 \caption {Particle distribution in impact parameter for the UA9 case.}
\end{figure}

\begin{figure}[b]
\hspace{-0.0cm} \vspace{0cm} {\includegraphics[width=8cm]
{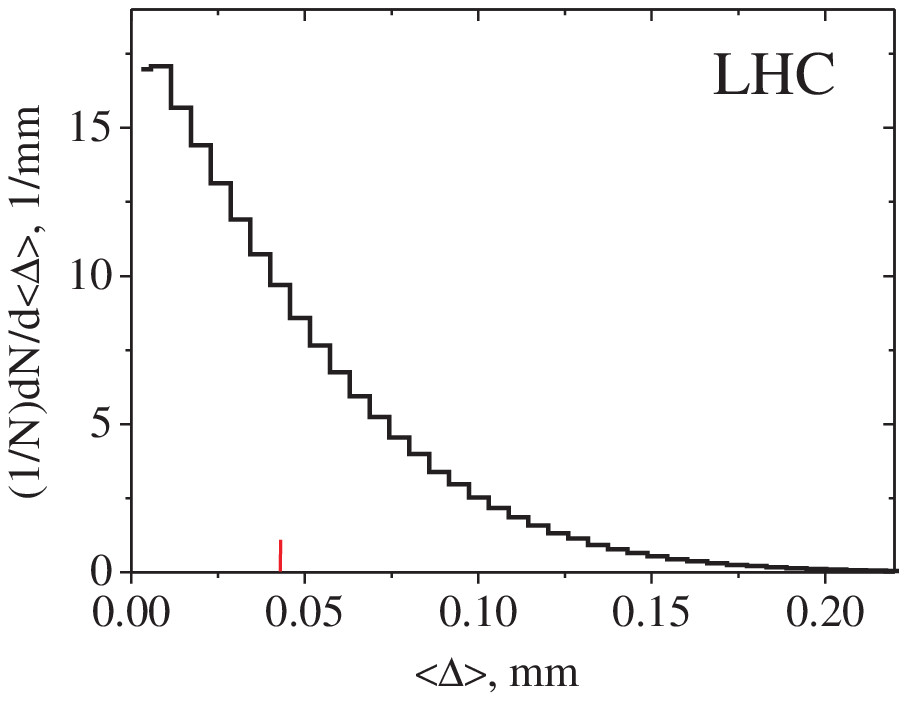}}
 \caption {The same for the LHC case.}
\end{figure}

In absence of phase correlations between betatron oscillations on
different revolution periods the phase can be sampled by the
formula $\psi = (2 \xi_2 -1)\pi$. A collision occurs with the
probability
\begin{equation} 
p_n = \frac{\psi(x_0(n))}{\pi} \simeq \frac{1}{\pi} \sqrt{\frac{2
n \delta}{x_c}}
\end{equation}
at the n-th revolution if
\begin{equation} 
|\psi| \leq \psi(x_0(n)).
\end{equation}

\begin{figure}[t]
\hspace{-0.0cm} \vspace{0cm} {\includegraphics[width=8cm]
{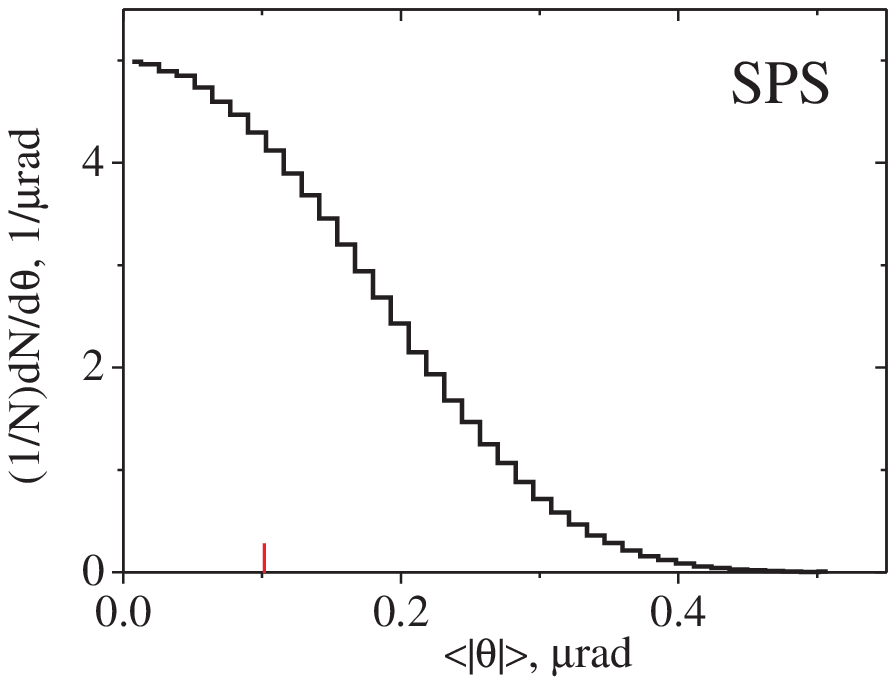}}
 \caption {Particle distribution in deflection angle for the UA9 case.}
\end{figure}

\begin{figure}[b]
\hspace{-0.0cm} \vspace{0cm} {\includegraphics[width=8cm]
{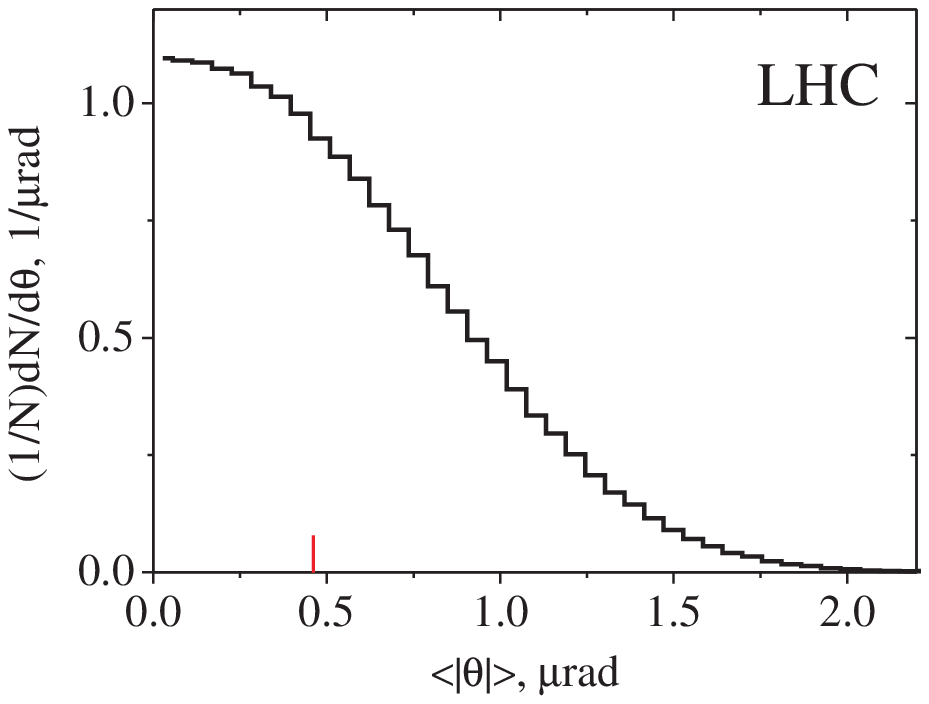}}
 \caption {The same for the LHC case.}
\end{figure}
If, otherwise, $|\psi| > \psi(x_0(n))$, no collision occurs and
one should continue the simulations with $n=n+1$ and so on. Since
the cumulative collision probability
\begin{equation} 
P_N = \sum_{n=1}^{n=N} p_n \simeq \frac{2}{3 \pi} \sqrt{\frac{2
\delta}{x_c}} N^{3/2}
\end{equation}
increases faster and faster with the revolution number $N$, the
"collision condition" (9) inevitably becomes fulfilled at some
revolution $N$ with some random values of $x_0(N)$ and $\psi$
allowing to evaluate both the collision coordinate (4) and angle
(6) of particle deflection at the moment of collision. Particle
distributions in the impact parameter $\Delta \equiv x-x_c$ and
deflection angle $\theta$, simulated for the UA9 and LHC cases,
are represented in Figs. 3-6. A three order in value difference in
the impact parameter in the UA9 and LHC cases is directly related
to that in the diffusion step -- see Table I.

Assuming that the collisions, naturally, occur at $P_N \sim 1$,
Eq. (10) can be reversed to estimate the typical revolution number
before the collision
\begin{equation} 
N \simeq \frac{(3 \pi P_N)^{2/3}}{2} \sqrt[3]{\frac{x_c}{\delta}}
\sim 2 \sqrt[3]{\frac{x_c}{\delta}},
\end{equation}
the average impact parameter
\begin{equation} 
\langle \Delta\rangle = P_N^{-1} \sum_{n=1}^{n=N} p_n
[x_0(n)cos(\psi)-x_c] \simeq \frac{3(3 \pi P'_N)^{2/3}}{10}
\sqrt[3]{x_c\delta^2} \sim 1.3 \sqrt[3]{x_c\delta^2}
\end{equation}
and absolute value of the deflection angle
\begin{equation} 
\langle |\theta|\rangle = P_N^{-1} \sum_{n=1}^{n=N} p_n
x_0(n)sin(\psi)/\beta \simeq \frac{3(3 \pi P''_N)^{1/3}}{8}
\frac{\sqrt[3]{x_c^2 \delta}}{\beta } \sim 0.8
\frac{\sqrt[3]{x_c^2 \delta}}{\beta }.
\end{equation}
\begin{figure}[t]
\hspace{-0.0cm} \vspace{0cm} {\includegraphics[width=9cm]
{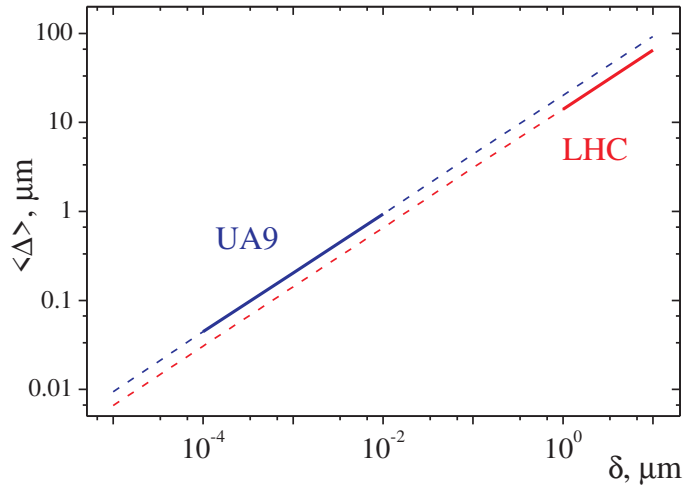}} \vspace{0cm}
 \caption {Average impact parameter {\it vs} average beam diffusion step for the SPS UA9 (upper curve) and the LHC (lower one).
 Solid parts mark the actual parameter regions.}\label{fig5}
\end{figure}
The right hand sides of Eqs. 11-13 contain numerical factors found
from the simulations and determining some effective collision
probabilities $P_N, P'_N$, $P''_N \sim 1$ which should be
considered as the parameters compensating a slight model
inconsistency. Despite the latter Eqs. 11-13 provide a sufficient
base to compare the conditions of the UA9 crystal collimation
experiment with the possible LHC crystal collimator performance as
well as to estimate the perspectives of possible crystal
collimation development \cite{tik1}. Fig. 7 again demonstrates
that the radical difference in diffusion steps results in the
drastic difference in average impact parameters in the UA9 and LHC
cases.

\section{Collimation efficiency for the UA9 experiments}

A simulated value of the impact parameter $\Delta$ can be directly
used to evaluate both the entrance transverse coordinate (4) and
angle (6), the knowledge of both of which is necessary to simulate
the particle trajectory inside the crystal in order to obtain the
angle of particle deflection by the latter. Recall that we measure
the angles from the direction of motion of the particle just
touching the collimator at its maximum displacement $x(\psi=0) =
x_0 = x_c$ from the beam axis -- see Fig. 2. Note {\it de bene
esse} that the chosen zero direction forms the angle
\begin{equation} 
\frac{dx(\psi=0)}{ds} =
\frac{1}{2}\sqrt{\frac{\varepsilon}{\beta}}\frac{d\beta}{ds} =
-\frac{x_c}{\beta}\alpha
\end{equation}
w.r.t. the beam axis. Here $s$ is the beam longitudinal coordinate
and $\alpha = -d\beta/ds/2$ is the conventional Twiss parameter.
In the ideal case a crystal has no miscut and its planes form zero
angle w.r.t. the chosen zero angle direction. Particles neither
enter the crystal through its lateral surface no leave the one
through it if they are channeled in this case.

The real situation is complicated by the inevitable presence of
both crystal miscut and crystal plane misalignment at the entrance
surface, characterized by the angles $\theta_m$ and $\theta_c$,
respectively. The crystal misalignment angle is assumed to be
positive if the planes are rotated in the direction of crystal
bending. If one determines the miscut angle as the one of crystal
plane rotation in the direction of crystal bending w.r.t. the
crystal lateral surface, the misorientation angle of the latter,
measured from the same zero angle direction, will be equal to
$\theta_{s0} = \theta_s(0) = \theta_c - \theta_m$. If the crystal
is bent with radius $R$, the surface tangential direction will
vary like $\theta_s(z) = \theta_{s0} + z/R$ with the longitudinal
coordinate $z \simeq s -s_c$. A behavior of the surface coordinate
\begin{equation} 
x_s(z) = x_s(0) + \int_0^z \theta_{s}(z)dz = x_s(0) + \theta_{s0}
z +z^2/2R,
\end{equation}
also measured in the crystal bending direction, considerably
differs if $\theta_{s0} >0$ and $\theta_{s0} < 0$ and, in the
latter case, if $|\theta_{s0}| > \varphi$ and $|\theta_{s0}| <
\varphi$, where $\varphi = l/R$ is the bending angle of the
crystal with length $l$. Namely, if $\theta_{s0} < 0$ a particle
can enter the crystal through the lateral surface and, if
$-\varphi < \theta_{s0} <0$, also leave it through the latter. On
the opposite, if $\theta_{s0} >0$ particles always enter the
crystal through the entrance face, while leave it either through
the lateral or exit ones. In all the cases the actual situation is
determined by the impact parameter $\Delta$, the random nature of
which allows for diverse trajectory types at any choice of
$\theta_s$ and $\varphi$. To get simple formulae for all possible
situations we first determined the minimal crystal surface
coordinate $x_{min} \equiv x_c$. After the Monte Carlo sampling of
the impact parameter value $\Delta = x(\psi)-x_c$ corresponding
transverse collision coordinate $x(\psi)$ was used to evaluate the
longitudinal one $z_{e}$ from the equation $x_s(z_{e}) = x(\psi)$.
Then both the particle entrance point coordinates $(x(\psi),
z_{e})$ and initial deflection angle (6) were used as the initial
conditions for its trajectory simulation inside the crystal. A
possibility to leave the crystal through the lateral surface at
some $z < l$ was permanently monitored using Eq. (15). The
particle transverse coordinate and deflection angle at the exit
together with the crystal position $s_c$ in the ring became the
initial conditions for the particle motion simulation in the
accelerator ring, for which the simplest model of betatron motion
was applied.

\begin{figure}[t]
\hspace{-0.0cm} \vspace{0cm} {\includegraphics[width=10cm]
{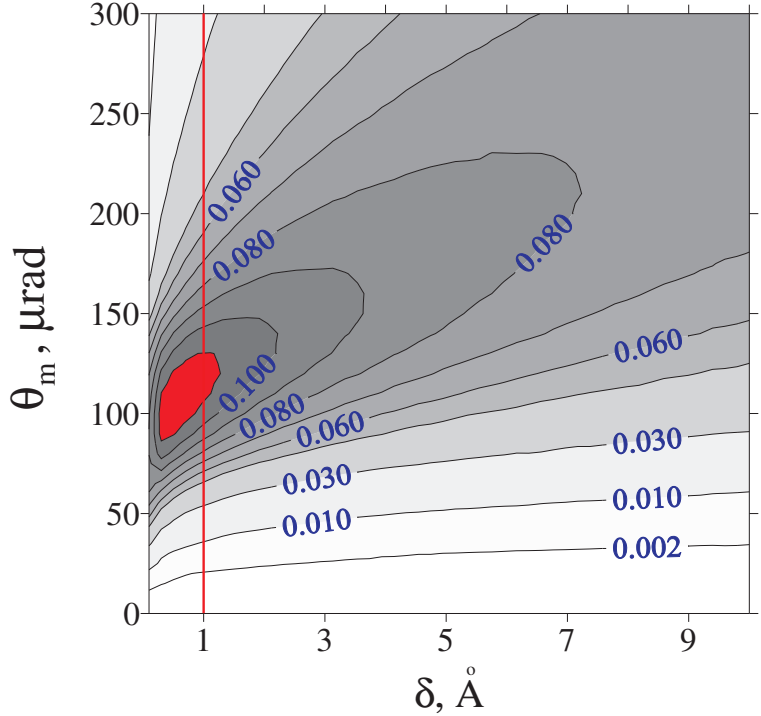}} \vspace{0cm}
 \caption {Measure in centimeters average length $\langle \Delta z \rangle$ of scattering of particles entering the
 crystal through the lateral crystal surface {\it vs} both miscut angle and diffusion step at perfect crystal plane alignment. }\label{fig6}
\end{figure}

However since the simulation of thousands of trajectories for tens
of different miscut angle and diffusion step values takes
considerable time, more rationally was first to use a simplified
fast approach allowing to elaborate a general view on the
influence of positive miscut on the collimation efficiency. The
idea originates from the mentioned proportionality of the decrease
of the channeling efficiency at the second crystal penetration to
the squared angle of multiple scattering of particles entering the
crystal first through the lateral surface. Since the latter, in
turn, is proportional to the scattering length $\Delta z$, one can
conclude that simply
\begin{equation} 
\Delta P_{ch} \propto \Delta z
\end{equation}
and reduce the issue to the analysis of the behavior of the
averaged length $\langle \Delta z \rangle$ of the first pass
through the crystal of the particles entering the latter
exclusively crossing its lateral surface. Fig. 8 illustrates the
simulated behavior of $\langle \Delta z \rangle$ in the typical
UA9 case of $l=2mm$ and $\varphi = l/R = 150 \mu rad$.
Surprisingly, the simulations unambiguously point to the region
$\theta_m \sim 100 \mu rad$ and $\delta \sim 1$ \AA~of the UA9
experiment parameters as to the one of the greatest possible
miscut influence on the collimation.

\begin{figure}[t]
\hspace{-0.0cm} \vspace{0cm} {\includegraphics[width=10cm]
{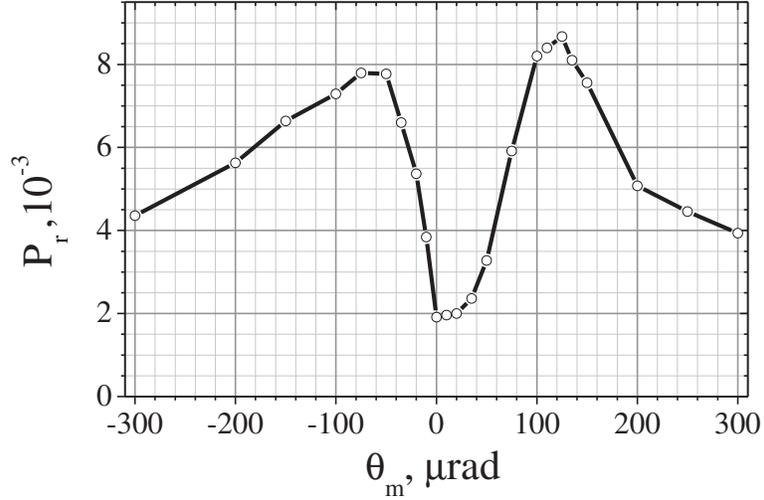}} \vspace{0cm}
 \caption {Probability of nuclear reactions in the crystal collimator {\it vs} miscut angle at perfect crystal plane alignment. }\label{fig7}
\end{figure}

Since the scattering length determines the decrease of the
channeling probability and since nonchanneled particles induce
more nuclear reactions that the channeled ones, Fig. 8 has to
simultaneously reflect the behavior of the rate of the nuclear
reactions induced in the crystal collimator. To illuminate the
possible role of positive miscut in the UA9 experiment we,
according to Ref. \cite{sca2} and Table I, had put $\delta = 1$
\AA~and conducted more detail Monte Carlo simulations of the
nuclear reactions in the miscut angle interval $-300 \mu rad \leq
\theta_m \leq 300 \mu rad$ taking now into detail consideration
also the particle motion in the crystal collimator. The dependence
obtained (see Fig. 9) demonstrates an evident agreement with that
of $\langle \Delta z \rangle$ along the vertical (red) line drawn
at $\delta = 1$ \AA~in Fig. 8, confirming thus the strong
influence of the positive miscut on the collimation process. In
principle, positive miscut causes even slightly larger increase in
nuclear reaction rate (the right peak) than the negative one (the
left one). At this the increase of the reaction probability caused
by the positive miscut with $\theta_m \simeq 125 \mu rad$ reaches
$8.6/1.9 \simeq 4.5$. Thus, the miscut influence, for sure, should
be taken into consideration for the full interpretation of the UA9
experiments \cite{sca}.

\section{Miscut influence at the LHC}

For the future possible application at the LHC it is important is
to clarify how the deteriorating miscut influence can be avoided.
A joint consideration of the particle motion in both the ring and
the crystal results in the encouraging conclusion that the
observed undesirable increase in nuclear reaction rate can be
easily avoided in both UA9 and LHC cases. In fact, some of the
conditions of the UA9 experiment prove to be practically optimal
for the demonstration of the maximum miscut role. The point is a
perfect matching of the average impact parameter (12) $\langle
\Delta \rangle \simeq 0.039 \mu m$ with the width $x_s(0)-x_c =
\theta_m^2 R/2 \simeq 0.067 \mu m$ of the impact parameter region
allowing particle entrance through the lateral crystal surface.
This matching made possible both the lateral entrance of the
majority of particles and their relatively continuous path inside
the crystal, the average value of which $\langle \Delta z \rangle
\simeq 1.2 mm$ exceeds a half of the crystal length $l=2mm$. It is
namely the nearly "amorphous" scattering at such a length which
gave the origin to the angular dispersion of particle beam causing
the decrease of the capture probability to the channeling regime
at their subsequent passages through the crystal collimator.

At least two ways to decrease the miscut role by fulfilling the
condition $\Delta \gg x_s(0)-x_c < \theta_m l$ can be readily
suggested. The most evident, though, probably, more difficult, is
to lessen the miscut angle down to about $10 \mu rad$, as Figs. 8
and 9 suggest. The second is to increase the collision parameter
(12) $\langle \Delta \rangle \propto \delta^{2/3}$ by means of
beam diffusion acceleration. While the diffusion step could be
nearly freely chosen in the collimation UA9 experiment, the actual
set of the LHC parameters solves this problem automatically.
Indeed, if $R = 100m$ and $l = 4 mm$, one obtains $x_s(0)-x_c
\simeq 0.32 \mu m$, or more than a {\it hundred times} less than
$\langle \Delta \rangle \simeq 43 \mu m \simeq 130(x_s(0)-x_c)$
without special measures. Thus, only a negligible portion of the
LHC protons will enter the crystal collimator through the lateral
crystal surface even at the typical miscut angles of $\theta_m
\sim 100 \mu rad$.

It also should be noted that despite the relatively large value of
the diffusion step $\delta$, the angular divergence (13) of the
colliding beam portion, as Fig. 6 demonstrates, is low enough to
provide the probability of capture into the regime of channeling
motion comparable to the maximum one. Nevertheless some decrease
in divergence remains desirable. The sharp dependence (3) of the
diffusion step on the collimator aperture $x_c/\sigma$ allows to
decrease $\delta$ by means of a slight decrease of the latter. At
this, if the divergence of the colliding beam portion is decreased
by several times, it will become possible to sharply rise the
probability of particle capture into the channeling regime up to
99\% by the method of the crystal cut \cite{tik1}.

In conclusion, the positive miscut influence indeed could increase
the nuclear reaction probability in the crystal collimator up to
4.5 times. Nevertheless if the crystal collimator system based on
the channeling particle deflection is realized at the LHC, its
functioning will not be considerably disturbed by the influence of
crystal miscut. In addition, the performance of the crystal
collimator can be drastically improved by the method \cite{tik1}
of the crystal cut.

\section{Acknowledgements }

One of the authors (V.T.) is obliged for an invitation to the UA9
Workshop to Dr. W. Scandale and Dr. G. Cavoto and also gratefully
acknowledges useful discussions with Prof. V. Guidi and Dr. A.
Mazzolari.


\end{document}